\documentclass{epl_ARC}

\usepackage{graphicx}
\usepackage{dcolumn}
\usepackage{bm}
\usepackage{psfrag}
\usepackage{amsmath,amssymb}

\title{Semilinear response}

\author{Michael Wilkinson\inst{1}, Bernhard Mehlig\inst{2} \and Doron Cohen\inst{3}}

\institute{\inst{1}Faculty of Mathematics and Computing, The Open
University, Walton Hall,
Milton Keynes, MK7 6AA, England \\
\inst{2}Department of Physics, G\"oteborg University, 41296
Gothenburg, Sweden \\
\inst{3}Department of Physics, Ben-Gurion University, Beer-Sheva,
84105, Israel }

\begin{document}

\maketitle

%%%%%%%%%%%%%%%%%%%%%%%%%%%%%%%%%%%%%%%%%%%%%%%%%%%%%%%%%%%%%%%%%%%%%%%
%%%%%%%%%%%%%%%%%%%%%%%%%%%%%%%%%%%%%%%%%%%%%%%%%%%%%%%%%%%%%%%%%%%%%%%

\begin{abstract}
We discuss the response of a quantum system to a time-dependent
perturbation with spectrum $\Phi(\omega)$. This is characterised
by a rate constant $D$ describing the diffusion of occupation
probability between levels. We calculate the transition rates by
first order perturbation theory, so that multiplying
$\Phi(\omega)$ by a constant $\lambda$ changes the diffusion
constant to $\lambda D$. However, we discuss circumstances where
this linearity does not extend to the function space of
intensities, so that if intensities $\Phi_i(\omega)$ yield
diffusion constants $D_i$, then the intensity $\sum_i
\Phi_i(\omega)$ does not result in a diffusion constant $\sum_i
D_i$. This \lq semilinear' response can occur in the absorption of
radiation by small metal particles.
\end{abstract}


%%%%%%%%%%%%%%%%%%%%%%%%%%%%%%%%%%%%%%%%%%%%%%%%%%%%%%%%%%%%%%%%%%%%%%%
%%%%%%%%%%%%%%%%%%%%%%%%%%%%%%%%%%%%%%%%%%%%%%%%%%%%%%%%%%%%%%%%%%%%%%%

1. {\sl Introduction}. We describe a previously unremarked
phenomenon concerning the response of a quantum system to a
time-dependent perturbation. The effect is significant if the
characteristic frequency scale of the perturbation $\omega_0$
obeys $\varrho\hbar\omega_0<1$, where $\varrho$ is the density of
states. Our analysis is an extension of linear-response theory, in
that it too relies on first-order perturbation theory, in our case
to derive rate constants for transitions between levels. The rate
constants are used in a master equation, which is then analysed
non-perturbatively. The response obtained is always linear in the
intensity of the driving perturbation, but in some circumstances
the response is not a linear functional of its spectrum. This
point will stated more precisely below. Conventional linear
response always describes the initial response of a system,
whereas our theory also considers how the response may differ
after an initial transient. The data plotted in figure \ref{fig:
1} (which is explained later) show that the predictions can differ
by orders of magnitude.

We characterise the response of the system by the rate $\dot
E\equiv{\rm d}E/{\rm d}t$ at which its energy is increased by the
action of the perturbation. This is directly related to
experimentally observable quantities, such as the absorption of
radiation by small metallic particles in an electromagnetic field.
The expectation from conventional linear-response theory (see for
example \cite{Kub62}) is that the rate of absorption is a linear
functional of the spectral intensity $\Phi(\omega)$ of the
radiation:
\begin{equation}
\label{eq: 1}
\dot{E}=\int_0^\infty {\rm d}\omega\ \alpha(\omega)\,
\Phi(\omega)
\end{equation}
where $\alpha(\omega)$ is a frequency-dependent absorption
coefficient. This expression satisfies two requirements
for linearity: if an intensity function $\Phi_i(\omega)$ results
in a response $\dot{E}_i$, then (\ref{eq: 1}) implies that for
some constant $\lambda$
\begin{eqnarray}
\label{eq: 2}
\Phi(\omega) \mapsto \lambda \Phi(\omega) \ \ \ \
& \ \ \ \Longrightarrow \ \ \ &
\dot{E}\mapsto \lambda \dot{E}\\
\label{eq: 3} \Phi(\omega) \mapsto  \sum_i \Phi_i(\omega)
& \ \ \ \Longrightarrow\ \ \ &
\dot{E}\mapsto \sum_i \dot{E}_i \ .
\end{eqnarray}
In this letter we introduce  a new form of linear-response theory, which
satisfies (\ref{eq: 2}) but not (\ref{eq: 3}), and which we
therefore term {\sl semilinear} response theory. In the limiting
case $\varrho\hbar\omega_0\ll 1$ we show that the absorption rate
is approximated by
\begin{equation}
\label{eq: 4} \dot{E}=\biggl[\int_0^\infty {\rm d}\omega\ \mu
(\omega)\, \Phi^{-1}(\omega) \biggr]^{-1}
\end{equation}
where $\mu (\omega)$ will be specified later. Equation (\ref{eq:
4}) clearly satisfies condition (\ref{eq: 2}), but not (\ref{eq:
3}). This is a consequence of the fact that (\ref{eq: 4}) is a
weighted harmonic average of the spectral intensity function,
$\Phi(\omega)$. Reference \cite{Coh05} discusses a model for a
quantum dot coupled to a conducting ring, where the DC conductance
is also obtained by a harmonic averaging procedure. In that case,
however, harmonic averaging is relevant because of the specific
structure of the Hamiltonian of that system. By contrast, the
results described here relating the AC response to $\Phi(\omega)$
are applicable to generic systems.

%%%%%%%%%%%%%%%%%%%%%%%%%%%%%%%%%%%%%%%%%%%%%%%%%%%%%%
\begin{figure}[bt]
\centerline{\includegraphics[width=10cm,clip]{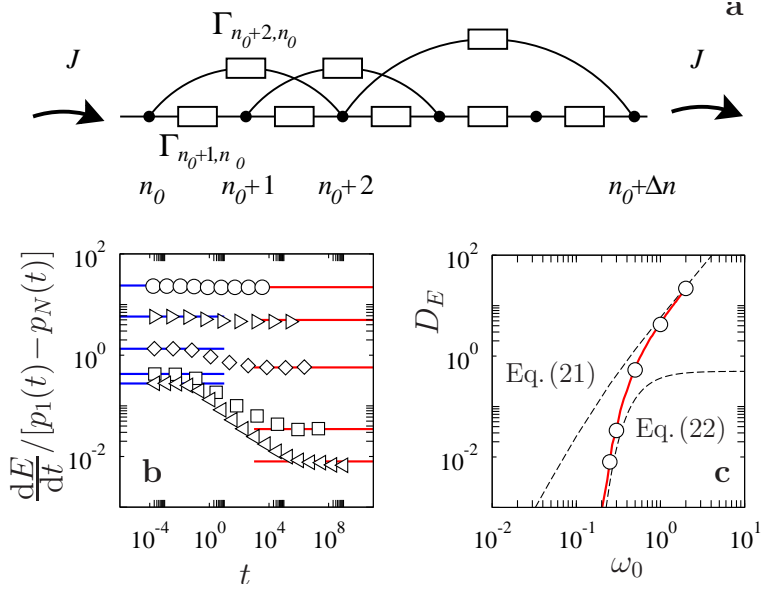}}
\caption{\label{fig: 1}
{\bf (a)}
The network of transitions between states
is analogous to a random resistor network.
{\bf (b)} Rate of energy absorption divided by difference in the
occupation probabilities of the lowest and the highest states,
showing crossover from an initial transient to a steady state. The
numerical simulations of (\ref{eq: 11}) employ random-matrix
eigenvalues with $\beta=1,\,\sigma^2=1,\,\varrho=1$, $N_{\rm p} =
3$, and $\Phi(\omega) =
\varepsilon^2\,\exp(-|\omega|/\omega_0)/\omega_0$, for
$\varepsilon^2 = 1$ and $\omega_0 = 2$ ($\circ$), $1$
($\triangleright$), $0.5$ ($\Diamond$), $0.3$ ($\Box$), and $0.25$
($\triangleleft$). The results were averaged over $200$
realisations of $\hat {\cal H}_j$. The asymptotes are: left,
linear-response for the initial transient [eq.(\ref{eq: 22})],
right, steady state determined by network model [eq.(\ref{eq:
14})].
{\bf (c)} Energy-diffusion constant $D_E$ vs. $\omega_0$.
Symbols are long-time asymptotes from figure \ref{fig: 1}{\bf b}.
The solid line is the semilinear response obtained by solving
(\ref{eq: 14}), averaging over $20$ realisations. Also shown are
linear-response approximation [eq.(\ref{eq: 22})], asymptote
to semilinear-response theory [eq.(\ref{eq: 23})]. }
\end{figure}
%%%%%%%%%%%%%%%%%%%%%%%%%%%%%%%%%%%%%%%%%%%%%%%%%%%%%%

Our approach is based upon an observation about the response of
quantum systems to low-frequency perturbations. We discuss a
system which absorbs a finite amount of energy, $E_0$, and
consider taking the limit as the characteristic frequency
$\omega_0$ of the perturbation approaches zero. The number of
quanta that the system absorbs, $E_0/\hbar \omega_0$, diverges as
$\omega_0\to 0$. In order to understand the response of a quantum
system to low-frequency perturbations, we must therefore consider
multiple excitations. We describe the excitation of the system by
a master equation, describing the probability $p_n(t)$ that the
system is in state with level number $n$ at time $t$. If this
master equation is treated in perturbation theory, we recover
conventional linear-response theory. However, our non-perturbative
treatment yields distinctive differences from the usual
linear-response results.

Our results are quite generally applicable, but it may be helpful
to bear in mind a specific example, namely a single electron
trapped inside an irregularly shaped enclosure, subjected to
fluctuating electric fields. This is a simplified model for the
absorption of electromagnetic radiation by small metallic
particles. In a classic paper, Gorkov and Eliashberg \cite{Gor65}
predicted a quantum size effect, where the absorption of radiation
would show distinctive structures for frequencies close to the
frequency $(\varrho\hbar)^{-1}$ (here $\varrho$ is the density of
states of single electron excitations at the Fermi energy). The
energy levels were assumed to have the same statistical properties
as random matrices, and the absorption was calculated using random
matrix models introduced by Dyson \cite{Dys62}, discussed in
\cite{Meh91}. We start from the same model and arrive at very
different conclusions. We remark that despite intensive
investigation, there is no clear experimental evidence for the
validity of the theory proposed in \cite{Gor65}.

%%%%%%%%%%%%%%%%%%%%%%%%%%%%%%%%%%%%%%%%%%%%%%%%%%%%%%%%%%%%%%%%%%%%%%%
%%%%%%%%%%%%%%%%%%%%%%%%%%%%%%%%%%%%%%%%%%%%%%%%%%%%%%%%%%%%%%%%%%%%%%%

2. {\sl The Hamiltonian}. We denote the Hamiltonian in the absence
of external fields by $\hat {\mathcal{H}}_0$. The perturbation is
described by a set of $N_{\rm p}$ operators $\hat {\mathcal{H}}_j$
multiplied by time-dependent fields $X_j(t)$:
\begin{equation}
\label{eq: 5} \hat {\mathcal{H}}(t)=\hat
{\mathcal{H}}_0+\sum_{j=1}^{N_{\rm p}} X_j(t)\hat {\mathcal{H}}_j
\ .
\end{equation}
In the case where the theory is applied to very small metal
particles in an electromagnetic field, $\hat {\mathcal{H}}_0$ is
the Hamiltonian for quasiparticle excitations, and the $\hat
{\mathcal{H}}_j$ are operators representing coupling of
quasiparticles to the $N_{\rm p}=3$ components of the electric
field, $X_j(t)$. The operators $\hat {\mathcal{H}}_j$ are not
simple dipole operators, because they must take account of
screening of the externally applied perturbation by polarisation
charges \cite{Str72,Aus93}.

The fields are not monochromatic, and their components $X_j(t)$
are random functions of time, satisfying
\begin{eqnarray}
\label{eq: 6} \langle X_j(t)\rangle{=}0\quad
\mbox{and}\quad
\langle X_i(t)X_{j}(t')\rangle{=}\delta_{ij}\phi(t{-}t')\,\,
\end{eqnarray}
(angular brackets denote averages). The fields have a spectral
intensity $\Phi(\omega)$, defined by
\begin{equation}
\label{eq: 7} \Phi(\omega)= \int_{-\infty}^\infty  {\rm d}\tau\
\phi(\tau) \exp({\rm i}\omega \tau)\,.
\end{equation}
In the numerical examples below, we take
$\Phi(\omega)=\varepsilon^2 \exp(-\vert
\omega\vert/\omega_0)/\omega_0$ where $\omega_0$, $\varepsilon$
are constants. An exponential dependence with $\omega_0=k_{\rm
B}T/\hbar$ is a natural choice if the system is excited thermally.

%%%%%%%%%%%%%%%%%%%%%%%%%%%%%%%%%%%%%%%%%%%%%%%%%%%%%%%%%%%%%%%%%%%%%%%
%%%%%%%%%%%%%%%%%%%%%%%%%%%%%%%%%%%%%%%%%%%%%%%%%%%%%%%%%%%%%%%%%%%%%%%

3. {\sl Master equation}. First consider the rate for transitions
between eigenstates of $\hat{\cal H}_0$. Expand the solution
$\vert \psi(t)\rangle$ of the Schr\" odinger equation ${\rm
i}\hbar{\rm d}\vert \psi\rangle/{\rm d}t=\hat
{\mathcal{H}}(t)\vert \psi \rangle$
\begin{equation}
\label{eq: 8} \vert \psi(t)\rangle =\sum_n a_n(t)\exp(-{\rm
i}E_nt/\hbar) \ \vert \varphi_n\rangle
\end{equation}
where $\hat{\mathcal{H}}_0\vert \varphi_n\rangle=E_n\vert
\varphi_n\rangle$ and the energies are ordered according to the
index $n$. The amplitudes $a_n(t)$ satisfy:
\begin{equation}
\label{eq: 9} \dot a_n={-{\rm i}\over{\hbar}}\sum_m \exp({\rm
i}\omega_{nm}t) \sum_{j=1}^{N_{\rm p}} \mathcal{H}_{nm}^{(j)}
X_j(t) a_m
\end{equation}
where $\mathcal{H}_{nm}^{(j)} = \langle \varphi_n\vert \hat
{\mathcal{H}}_j\vert \varphi_m\rangle$ and
$\omega_{nm}=(E_n-E_m)/\hbar$. Solving perturbatively for the
initial condition $a_n(0)=\delta_{nm}$ one obtains an expression
for $a_n(t)$ and hence for the probability $p_n(t)=\langle \vert
a_n(t)\vert^2 \rangle$ to be in the $n^{\rm th}$ state. For $n \ne
m$ we have $p_n(t)=\Gamma_{nm}t+O(t^2)$, where the rate constants
are given by a version of Fermi's golden rule
\begin{equation}
\label{eq: 10} \Gamma_{nm} \ \ = \ \ \frac{1}{\hbar^2} \
\Phi(\omega_{nm}) \ \sum_{j=1}^{N_{\rm p}} \vert
\mathcal{H}_{nm}^{(j)}\vert^2\,.
\end{equation}
The expression for $p_n$ is valid for times sufficiently short
that $1-p_m \ll 1$, but large enough that $\omega_0 t\gg 1$. These
conditions are compatible for a sufficiently small $\varepsilon$.

The rate constants can be used to write the master equation for
the occupation probabilities $p_n(t)$:
\begin{equation}
\label{eq: 11} {{\rm d}p_n\over{{\rm d}t}}
=\sum_{m}\Gamma_{nm}(p_m-p_n)\,.
\end{equation}
This master-equation ignores interference effects,
which average away on long timescales.

Our model (\ref{eq: 5}), (\ref{eq: 6}) describes excitation of a
quantum system due to a perturbation. In practical applications,
the system may also be subject to relaxation effects. The
master-equation model can then be augmented with terms
representing relaxation processes. These additional terms could
represent the transfer of energy from electronic excitations into
phonons or photons. Electron-electron interactions could also be
included, although these represent re-arrangement of energy within
the electronic system rather than relaxation. The lifetime for an
electron to emit photons or phonons diverges as the energy of
excitation of the electron approaches zero. For systems excited by
low-frequency fields, electrons may therefore be excited by many
quanta before their relaxation rate is significant. Thus our
approach is in contrast with conventional linear response theory
\cite{Kam95} which implicitly assumes that multiple excitations
are not relevant.

In the case of absorption of electromagnetic radiation by small
conducting particles, reference \cite{Gor65} shows that phonons do
not cause relaxation at low frequencies, so that emission of
photons is the dominant relaxation mechanism. In this case it is
easy to see that there is multiple excitation when the intensity
of radiation at frequency $\omega_0$ is large compared to the
intensity of black-body radiation at temperature
$T=\hbar\omega_0/k_{\rm B}$. This condition is easily satisfied at
the microwave or far-infrared frequencies which are relevant to
experimental studies on the effect discussed in \cite{Gor65}.

%%%%%%%%%%%%%%%%%%%%%%%%%%%%%%%%%%%%%%%%%%%%%%%%%%%%%%%%%%%%%%%%%%%%%%%
%%%%%%%%%%%%%%%%%%%%%%%%%%%%%%%%%%%%%%%%%%%%%%%%%%%%%%%%%%%%%%%%%%%%%%%

4. {\sl Energy diffusion and resistor networks}. We are interested
in the long-time behaviour of the master equation (\ref{eq: 11}).
The coarse-grained occupation probabilities obey a continuity
equation:
\begin{equation}
\label{eq: 12} \partial_t p(n,t) + \partial_n J(n,t) = 0
\end{equation}
with probability flux $J$.
We argue below that the coarse-grained
occupation probability obeys
Fick's law, ${J=-D\partial p/\partial n}$,
so that $p(n,t)$ obeys a diffusion equation
\begin{equation}
\label{eq: 13} \partial_t p
 = \partial_n \left[ D \partial_n p \right]\,.
\end{equation}
In order to determine $D$ we make use of the analogy between
(\ref{eq: 11}) and Kirchoff's equation for a resistor network
(illustrated in figure \ref{fig: 1}{\bf a}): nodes $n$ and $m$
are connected by conductances $G_{nm}$. If a
current $I_n$ is supplied at node $n$, the potentials $V_n$
satisfy
\begin{equation}
\label{eq: 14} I_n=\sum_m G_{nm}(V_n-V_m)\,.
\end{equation}
The probabilities $p_n$ in (\ref{eq: 11}) correspond to the
potentials $V_n$, the rates $\Gamma_{nm}$ to the conductances
$G_{nm}$, and in the steady state $I_n \equiv -{\rm d}p_n/{\rm d}
t=0$  at all the nodes. Coarse graining is achieved by considering
a truncated network segment of length $\Delta n \gg 1$ as
illustrated in figure~\ref{fig: 1}{\bf a}, with a current $J$ is
injected into one end and extracted at the other end. This finite
segment is described by an equation in the form of (\ref{eq: 14})
with ${I_n=J\delta_{n,n_0}-J\delta_{n,n_0+\Delta n}}$, resulting
in a potential difference $\Delta V=V_{n_0+\Delta n}-V_{n_0}$. We
expect that $\Delta V/\Delta n$ approaches a limit as $\Delta n$
increases, implying a `coarse grained' Fick's law: ${J\approx
-D(p_{n_0{+}\Delta n}-p_{n_0})/\Delta n}$, which is analogous to
Ohm's Law. Thus $D/\Delta n$ is the conductance of the segment,
and $D$ is obtained as $D=-J\lim_{\Delta n\to \infty}{\Delta
n/\Delta V}$. We remark that Miller and Abrahams \cite {Mil60}
introduced random resistor models in studies of spatial (as
opposed to energy) diffusion in disordered systems.

In what follows we discuss expressions for $D$ in two limiting
cases. When the rate constants are negligible for all but
nearest-neighbour transitions, the resistance $(D/\Delta n)^{-1}$ is the
sum of resistors in series, leading to
\begin{equation}
\label{eq: 15} D = \bigg(\lim_{\Delta n\to\infty} \frac{1}{\Delta
n} \sum_{n=n_0}^{n_0{+}\Delta n}\Gamma_{n,n+1}^{-1} \bigg)^{-1}
\equiv \left\langle \Gamma_{n,n+1}^{-1}\right\rangle^{-1}
\end{equation}
To show that Fick's law holds in the limit considered here, it
suffices to show that (\ref{eq: 15}) yields a finite result for
$D$ (an example is given in equation (\ref{eq: 23}) below). It is
an immediate consequence of the nature of the harmonic average
that the diffusion constant is significantly reduced by the
presence of \lq bottlenecks', that is links with very low
transition rates \cite{Ale81}. Reference \cite{Coh05} considers
the consequences of this for energy diffusion in a model for the
DC response of a quantum dot coupled to a conducting ring.

We turn to the other extreme case, where many transitions have
significant weight (not just to near neighbours).  We may assume
that the potential changes linearly along the network ($p_n\propto
n$) and find that
\begin{equation}
\label{eq: 16} D = \big \langle \ {\textstyle{1\over 2}}  \sum_{m}
(m-n)^2 \Gamma_{n,m} \big \rangle \,.
\end{equation}
where the angular brackets means averaging over $n$ as in equation
(\ref{eq: 15}). To derive this, note that the contribution of bond
of length $\Delta n = (m-n)$ to the current is proportional to the
potential drop and hence to $\Delta n$. Furthermore the number of
bonds of length $\Delta n$ passing through a given section gives a
further factor $\Delta n/2$.

%%%%%%%%%%%%%%%%%%%%%%%%%%%%%%%%%%%%%%%%%%%%%%%%%%%%%%%%%%%%%%%%%%%%%%%
%%%%%%%%%%%%%%%%%%%%%%%%%%%%%%%%%%%%%%%%%%%%%%%%%%%%%%%%%%%%%%%%%%%%%%%

5. {\sl Rate of absorption of energy}. The expectation value of
the energy of the system is
\begin{equation}
\label{eq: 17} E(t)=\sum_n p_n(t) E_n\,.
\end{equation}
When many states are
excited, the sum may be approximated by an integral. Applying
(\ref{eq: 13}), the rate at which energy is absorbed is
\begin{equation}
\label{eq: 18} \dot{E} \!=\! \int\!\! {\rm d}n\, E_n
\partial_n [D \partial_n p]
=\!-\!\int\!\! {\rm d}E\, \varrho D_E{\partial p\over{\partial
E}}\,.
\end{equation}
This equation shows that the rate of energy absorption $\dot{E}$
is proportional to the energy-diffusion coefficient ${D_E =
D/\varrho^2}$, a principle that was introduced in \cite{Wil88}.

%%%%%%%%%%%%%%%%%%%%%%%%%%%%%%%%%%%%%%%%%%%%%%%%%%%%%%%%%%%%%%%%%%%%%%%
%%%%%%%%%%%%%%%%%%%%%%%%%%%%%%%%%%%%%%%%%%%%%%%%%%%%%%%%%%%%%%%%%%%%%%%

6. {\sl An example of semilinear response}. Consider the case
where $\varrho \hbar\omega_0 \ll 1$. Here the rate constants
(\ref{eq: 10}) decrease very rapidly as the separation in energy
increases, and we can neglect all of the rate constants
$\Gamma_{nm}$ other than those describing nearest neighbour
coupling. The diffusion constant ${D_E= D/\varrho^2}$ is then
estimated via equation (\ref{eq: 15}). It is clear that large gaps
in the spectrum create `bottlenecks' which slow the diffusion of
probability. We define $P(S){\rm d}S$ as the probability that the
normalised spacing between two successive levels
$(E_{n+1}-E_n)\varrho$ is in the interval $[S,S+{\rm d}S]$. We
assume that the matrix elements $\mathcal{H}_{nm}^{(j)}$ are
independent Gaussian random variables with variance $\sigma^2$ and
zero mean, independent of the energy levels. We write
$\mathcal{H}_{nm}^{(j)} = \sigma x_{nm}^{(j)}$ with Gaussian
random variables $x_{nm}^{(j)}$, each with zero mean and unit
variance, then substitute (\ref{eq: 10}) into (\ref{eq: 15}) and
find
\begin{equation}
\label{eq: 20} D_E\!=\!{\sigma^2\over {(\varrho\hbar)^3}} \biggl[
\int\!\! \frac{{\rm d}{\bf x}\,{\rm e}^{-{\bf
x}^2/2}}{(2\pi)^{N_{\rm p}/2}{\bf x}^2} \biggr]^{-1}
\biggl[\int_0^\infty\!\!\!\!\!\! {\rm d}\omega
\frac{P(\varrho\hbar\omega)}{ \Phi(\omega)} \biggr]^{-1}\,.
\end{equation}
Eq. (\ref{eq: 20}) is an example of semilinear response, in the
form of eq.(\ref{eq: 4}). Assuming that $\Phi(\omega)$ decreases
rapidly when $\omega\gg \omega_0$, the integral in (\ref{eq: 20})
is dominated by the tail of the level spacing distribution.
Denoting the second term (with the integral over ${\rm d}{\bf x}$)
by $W$, we find that $W=1$ for $N_{\rm p}=3$. For $N_{\rm p} =
1,2$ we find $W=0$, so that $D_E=0$, implying that the spread of
probability is sub-diffusive.

%%%%%%%%%%%%%%%%%%%%%%%%%%%%%%%%%%%%%%%%%%%%%%%%%%%%%%%%%%%%%%%%%%%%%%%
%%%%%%%%%%%%%%%%%%%%%%%%%%%%%%%%%%%%%%%%%%%%%%%%%%%%%%%%%%%%%%%%%%%%%%%

7. {\sl Linear-response theory}. Now we contrast (\ref{eq: 20})
with conventional linear-response theory. We assume that the
initial probability $p_n(0)$ is a smooth function of the energy of
the state. We differentiate (\ref{eq: 11}), substitute in
(\ref{eq: 17}), and expand $p_m-p_n$ to first order
in $E_n-E_m$. Interchanging the indices $n$ and $m$ and
averaging the two expressions for the double sum gives:
\begin{equation}
\label{eq: 21} \dot{E} = -{\textstyle{\frac{1}{2}}} \sum_{n,m}
(E_m-E_n)^2 \Gamma_{nm} \frac{\partial p}{\partial E}\,.
\end{equation}
We can identify $D_E$ by comparison with (\ref{eq: 18}). After
substitution of (\ref{eq: 10}) we obtain an expression in terms of
the two-level correlation function
$R_2(\epsilon)=\sum_{nm}\langle\delta(E-E_n)\delta(E+\epsilon-E_m)\rangle/\varrho^2$
(we use the notation of \cite{Meh91})
\begin{equation}
\label{eq: 22} D_E=
N_{\rm p}\sigma^2\hbar\varrho\int_0^\infty\!\!\! {\rm d}\omega
\,\omega^2\, R_2(\hbar\omega)\Phi(\omega)\,.
\end{equation}
This a linear functional of the spectral intensity $\Phi(\omega)$,
leading to and expression of the form (\ref{eq: 1}). Equation
(\ref{eq: 22})  is a version of the \lq Kubo formula' of
linear-response theory and is equivalent to a result obtained by
Gorkov and Eliashberg \cite{Gor65}. It is subject to the criticism
that it only describes the initial response of the system (figure
\ref{fig: 1}{\bf b}): after a short transient, the probabilities
may cease to be given accurately by a smooth function of the
energy, leading to a very different rate of absorption, such as
that given by (\ref{eq: 20}). Conventional linear-response theory
\cite{Kam95} implicitly assumes that strong relaxation prevents
level-number diffusion from exploring the \lq bottlenecks'.

Finally we remark that if  the
initial probability $p_n(0)$ is a smooth of level number
$n$ instead of $E_n$, a slightly modified form
of linear-response theory is obtained \cite{Kam95}.

%%%%%%%%%%%%%%%%%%%%%%%%%%%%%%%%%%%%%%%%%%%%%%%%%%%%%%%%%%%%%%%%%%%%%%%
%%%%%%%%%%%%%%%%%%%%%%%%%%%%%%%%%%%%%%%%%%%%%%%%%%%%%%%%%%%%%%%%%%%%%%%

8. {\sl Random-matrix models and numerical experiments}. We now
compare the calculation of $D_E$ using (\ref{eq: 20}) with
conventional linear-response theory (\ref{eq: 22}), assuming
random matrix models for $P(S)$ and $R_2(\epsilon)$. Although the
spectra of complex quantum systems differ, their statistical
properties are very similar and can be calculated for suitably
defined random-matrix ensembles \cite{Meh91}. There are three \lq
universal' ensembles, labelled by an integer index $\beta\in
\{1,2,4\}$. For $P(S)$ we use the \lq Wigner surmise', $P(S)\sim
a_\beta S^\beta \exp(-c_\beta S^2)$, with $a_\beta$ and $c_\beta$
chosen so that $P(S)$ is normalised with mean value unity. The
diffusion coefficient (\ref{eq: 20}) depends on the large
separations, so that we require accurate information about the
values of $P(S)$ for large argument: precise information about the
large $S$ asymptotics is given in \cite{Meh91}, but for the
moderately large values of $S$ that are probed by our numerical
studies, the Wigner surmise gives more accurate results. We also
require $R_2(\epsilon)$, to evaluate (\ref{eq: 22}). Here it is
the behaviour for small spacings that is of most interest, where
$R_2(\epsilon)=k_\beta
(\varrho\epsilon)^\beta+O(\epsilon^{\beta+1})$ with universal
constants $k_\beta$.

We illustrate the theory by comparing the predictions of (\ref{eq:
20}) and (\ref{eq: 22}) for a spectral intensity
$\Phi(\omega)=\varepsilon^2\exp(-\vert \omega\vert
/\omega_0)/\omega_0$ and $\beta =1$, using the Wigner surmise for
$P(S)$. As $\omega_0\to 0$, equation (\ref{eq: 20}) predicts that
$D_E$ approaches zero in a non-analytic fashion:
\begin{equation}
\label{eq: 23} D_E=(\varepsilon^2\sigma^2/{ 2\varrho\hbar})\,
\exp\{-1/[\pi(\varrho \hbar \omega_0)^2]\}\,.
\end{equation}
This is dramatically different from the result of conventional
linear response theory (\ref{eq: 22}) where, for small values of
$\omega_0$, we find $D_E \sim  C_\beta N_{\rm p} \sigma^2
\varepsilon^2 (\hbar \varrho)^{\beta+1} \omega_0^{\beta+2}$ (for
some universal constants~$C_\beta$).

Figure \ref{fig: 1}{\bf b},{\bf c} shows numerical results for
simulations using random-matrix energy levels, with $\beta=1$,
$\varepsilon=1$, $\sigma^2 =1$, $N_{\rm p}=3$, and $\varrho=1$.
The data in figure \ref{fig: 1}{\bf b} are obtained by a
simulation of (\ref{eq: 11}) with (\ref{eq: 17}). We use GOE
\cite{Dys62} random matrices of dimension $N=4000$ and an
exponential initial distribution $p_n\propto\exp(-E_n/\Delta E)$
with $\varrho \Delta E=100$. After an initial transient both
$\dot{E}$ and $(p_1(t)-p_N(t))$ decrease, their ratio approaching
a limit which (using (\ref{eq: 18})) we identify as $\varrho D_E$.
These limiting values of $D_E$ are plotted as symbols in figure
\ref{fig: 1}{\bf c}. The data for the resistor network (solid
line) was obtained by solving Kirchoff's law for a network with
$N=4000$ nodes, using singular-value decomposition.

%%%%%%%%%%%%%%%%%%%%%%%%%%%%%%%%%%%%%%%%%%%%%%%%%%%%%%%%%%%%%%%%%%%%%%%
%%%%%%%%%%%%%%%%%%%%%%%%%%%%%%%%%%%%%%%%%%%%%%%%%%%%%%%%%%%%%%%%%%%%%%%

9. {\sl Summary}. We have considered the non-perturbative solution
of a master equation describing transitions between levels. Its
solutions are in general diffusive for large times, with a
diffusion constant obtained from the conductivity of a random
resistor network. When the characteristic frequency $\omega_0$ is
small ($\varrho \hbar \omega_0\ll 1$), only transitions between
neighbouring levels are significant, analogous to resistors in
series. The diffusion constant is then the harmonic mean of the
rate constants. It is determined by the tail of the level spacing
distribution and is an example of semilinear response.

%%%%%%%%%%%%%%%%%%%%%%%%%%%%%%%%%%%%%%%%%%%%%%%%%%%%%%%%%%%%%%%%%%%%%
%%%%%%%%%%%%%%%%%%%%%%%%%%%%%%%%%%%%%%%%%%%%%%%%%%%%%%%%%%%%%%%%%%%%%

10. {\sl Acknowledgements}. We thank Y. Gefen for helpful
discussions. This research was supported by the Vetenskapsr\aa det
and by the Israel Science Foundation (grant No.11/02).

%%%%%%%%%%%%%%%%%%%%%%%%%%%%%%%%%%%%%%%%%%%%%%%%%%%%%%%%%%%%%%%%%%%%%
%%%%%%%%%%%%%%%%%%%%%%%%%%%%%%%%%%%%%%%%%%%%%%%%%%%%%%%%%%%%%%%%%%%%%

%%%%%%%%%%%%%%%%%%%%%%%%%%%%%%%%%%%%%%%%%%%%%%%%%%%%%%%%%%%%%%%%%%%%%
%%%%%%%%%%%%%%%%%%%%%%%%%%%%%%%%%%%%%%%%%%%%%%%%%%%%%%%%%%%%%%%%%%%%%
\end{document}